\documentclass{mnras}
\usepackage{mathtools}
\pdfoutput =1

\usepackage{amssymb}
\usepackage{graphicx}
\usepackage{caption}
\usepackage{amsmath}

\usepackage{comment}

\pubyear{2019}
\title[Observing Flare Stars Below 100 MHz with the LWA]{Observing Flare Stars Below 100 MHz with the LWA}
\author[I. Davis et al.]{
Ivey Davis,$^{1}$
Greg Taylor,$^{1}$ 
Jayce Dowell$^{1}$ \\
$^{1}$Department of Physics and Astronomy, University of New Mexico, 210 Yale Blvd NE, Albuquerque, New Mexico
}
\begin{document}
\pagerange{\pageref{firstpage}--\pageref{lastpage}}
\label{firstpage}

\maketitle

\begin{abstract}
     We observed the flare stars AD Leonis, Wolf 424, EQ Pegasi, EV Lacertae, and UV Ceti for nearly 135 hours.  These stars were observed between 63 and 83 MHz using the interferometry mode of the Long Wavelength Array.  Given that emission from flare stars is typically circularly polarized, we used the condition that any significant detection present in Stokes I must also be present in Stokes V at the same time in order for us to consider it a possible flare. Following this, we made one marginal flare detection for the star EQ Pegasi. This flare had a flux density of 5.91 Jy in Stokes I and 5.13 Jy in Stokes V, corresponding to a brightness temperature $1.75 \times 10^{16}(r/r_*)^{-2}$ K.
     \newline
\end{abstract}

\section{Introduction}
Flares are not an uncommon event for magnetically active stars.  For F and G type stars, there is a striking resemblance between these flares and solar flares, including the presence of both compact and two-ribbon flares that conveniently allows for the implementation of a solar analogy for understanding the properties of these flares (Pallavicini, Tagliaferri \& Stella 1990). Given the dependence of flaring on magnetic field dynamics, particularly on magnetic reconnection, it makes sense that M-dwarfs, which have been found to have magnetic field strengths on the order of kilo Gauss (Johns-Krull \& Valenti 1996), would be particularly active.\par
One property of M-dwarfs that is of interest due to its probable role in the strength of the magnetic field is that they are entirely or near entirely convective (Reiners \& Basri 2009).  This characteristic means that stars with masses less than 0.35\(M_\odot\) will lack the interface between the convective and radiative layer that is related to the efficiency of magnetic field generation in solar-type stars (Charbonneau 2005).  The interior dynamics may also affect the kind of dynamo effect involved in generating the magnetic field; while the Sun's field is heavily influenced by the $\alpha$-$\omega$ dynamo, fully convective stars are more likely to have a field driven by convective turbulence (Durney, Young, \& Roxburgh 1992; Yang \textit{et al.} 2017). It has also been observed that the rotation period of the star is related to the magnetic field strength, so the rapid rotation of these stars likely also contributes to these magnetic fields with strengths thousands of times that of the solar magnetic field (G{\"u}del 2002). \par
Radio emission of flare events has two main variants--- coherent and incoherent (Benz \& G{\"u}del 2010).  The cause of this emission is still being investigated; while some claim that electron cyclotron maser emission (ECM) is involved (Bastian \& Bookbinder 1987; Abdul-Aziz \textit{et al.} 1995), the work of others indicate that ECM emission cannot be supported by their results (Abranin \textit{et al.} 1998).  Important factors which help discriminate the type of emission are the polarization and the duration of the flare.  Emission is either strongly circularly polarized and short in duration or exhibits little to no polarization and is long lasting.  The former suggests coherent emission like ECM or fundamental plasma emission and the latter is characteristic of gyrosynchrotron emission. In the circumstance of coherent emission, further details about the emission can be derived through the spectrum-- since emission from a single maser is inherently narrowband, coherent, broadband emission implies either plasma emission or multiple maser sites being active.\par
The difficulty in understanding the emission process of flare stars, particularly M-dwarfs, is driven by the inconclusive results at frequencies below 1 GHz.  Even within individual papers, results both support and oppose the solar analogy for stellar flares.  There has been evidence of both negative frequency drifts -- a feature of type II solar flares-- as well as positive frequency drifts.  Observations by Boiko, Konovalenko, Koliadin, \& Melnik (2012) indicate frequent flaring at frequencies as low as 25 MHz, while Jackson, Kundu \& Kassim (1990) and Lynch \textit{et al.} (2017) only find marginal possible detections at 154 MHz. In order to better understand the origin of these discrepancies, we observed five nearby flare stars between 63 and 83 MHz.
\section{Methods}
\begin{table*}
    \small
    \centering
    \begin{tabular}{c|c|c|c|c}
        \hline \hline
        \textbf{Star} & \textbf{Distance [pc]} & \textbf{Spectral Type} & \textbf{RA (J2000)} & \textbf{Dec (J2000)}   \\ \hline 
         AD Leo & 4.9 & M3.5eV& 10 19 35.63 & 19 52 11.34 \\ 
         Wolf 424 & 4.3 & 	dM6e/dM6e & 12 33 15.22 & 09 01 19.42\\
         EQ Peg &  6.4 & M3.5V/M4.5V & 23 31 52.91 & 19 56 13.04 \\ 
         EV Lac & 5.02 & M3.5 & 22 46 48.51 & 44 19 54\\
         UV Ceti & 2.68 & M5.5V/M6V & 01 39 05.65& -17 56 50\\\hline \hline
    \end{tabular}
    \caption{Information on the flare star or flare star systems observed. The RA and DEC provided here account for proper motion. The two values of Spectral Type given for Wolf 424, EQ Peg, and UV Ceti are due to the fact that they are binary systems.  The coordinates and spectral type were provided by SIMBAD and the distances by Bastian (1990).} %proper motion should be taken into account for EQ Peg and Wolf.
    \label{tab:flarestarinfo}
\end{table*}
The five flare stars observed were AD Leonis (AD Leo), Wolf 424, EQ Pegasi (EQ Peg), EV Lacertae (EV Lac), and UV Ceti.  Information on these stars is provided in Table \ref{tab:flarestarinfo}. All of these stars are UV Ceti type variables that have been detected at  frequencies at least as low as $\sim$1 GHz, are no farther than 10 pc away, and had transit elevations of at least 45$^{\circ}$ at the Long Wavelength Array (LWA) sites (a declination of $-10^{\circ}$).  The exception to this was UV Ceti, which had a transit elevation of about 37$^{\circ}$.  \par
Observations of the flare stars were taken with the LWA, a dipole array with 256 elements at each of the sites.  In order to distinguish the sources in the sky, the interferometry mode was used, employing the site near the Very Large Array (LWA1) and the site at the Sevilleta National Wildlife Refuge (LWA-SV), giving a $\sim$70 km baseline. Such a separation aided not only in angular resolution, but also in mitigation of locale-specific radio frequency interference (RFI) during the correlation process. Two beams operated simultaneously at each site, allowing for simultaneous observing of calibrators and the source.  One beam per site alternated between a flux and phase calibrator (calibration beam) and the other beam alternated between a check source-- which served to confirm whether the beam was operating properly-- and the flare star of interest (source beam).  The phase calibrators chosen were within a few degrees of the respective star in order to account for ionospheric effects in the region of the star.  Information on the phase and flux calibrators and the check sources are found in Table \ref{tab:obs_info}. \par
\begin{figure}
    \centering
    \begin{tabular}{cc}
        \includegraphics[width=0.36\textwidth]{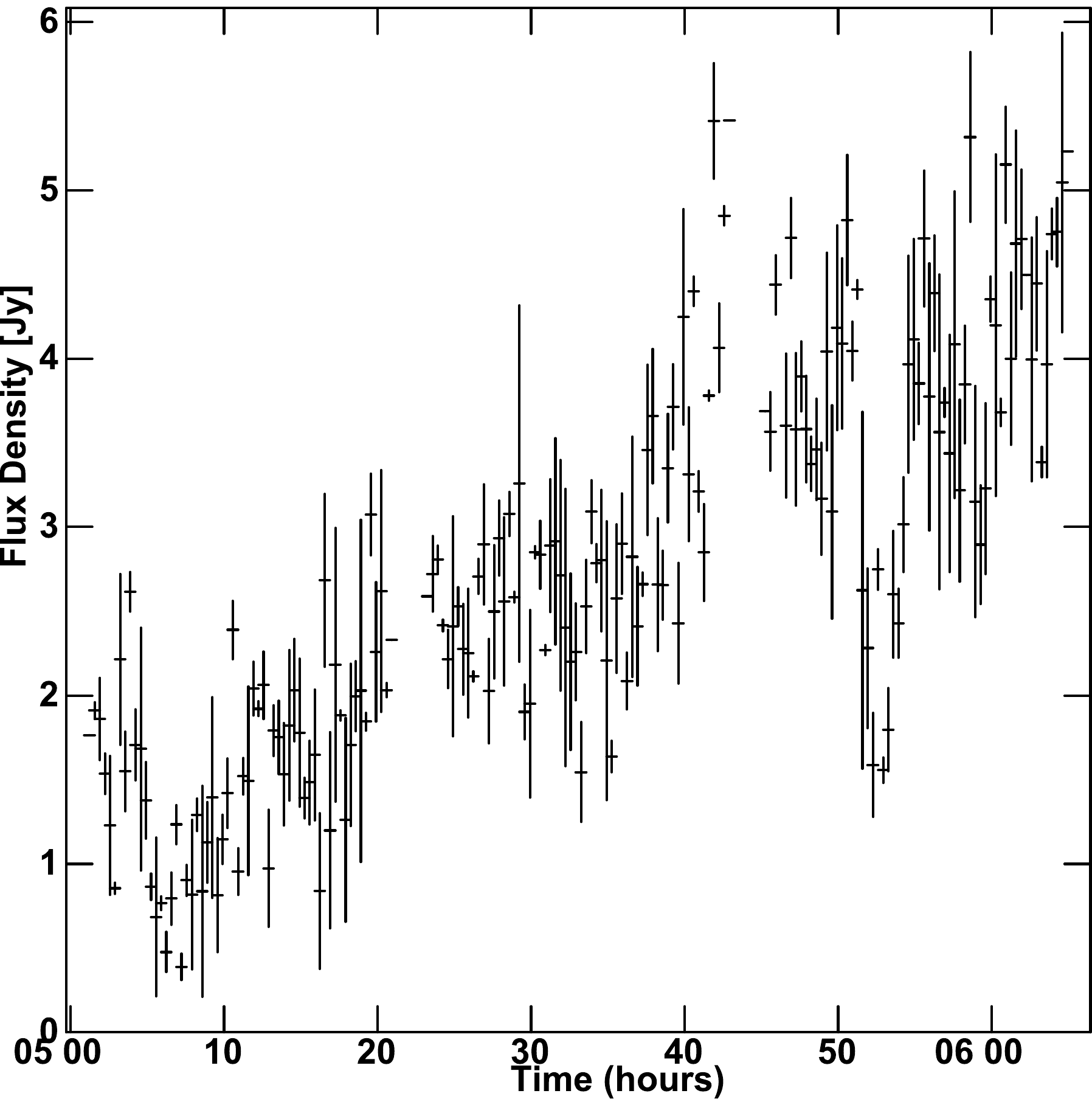} \\ 
        \includegraphics[width=0.36\textwidth]{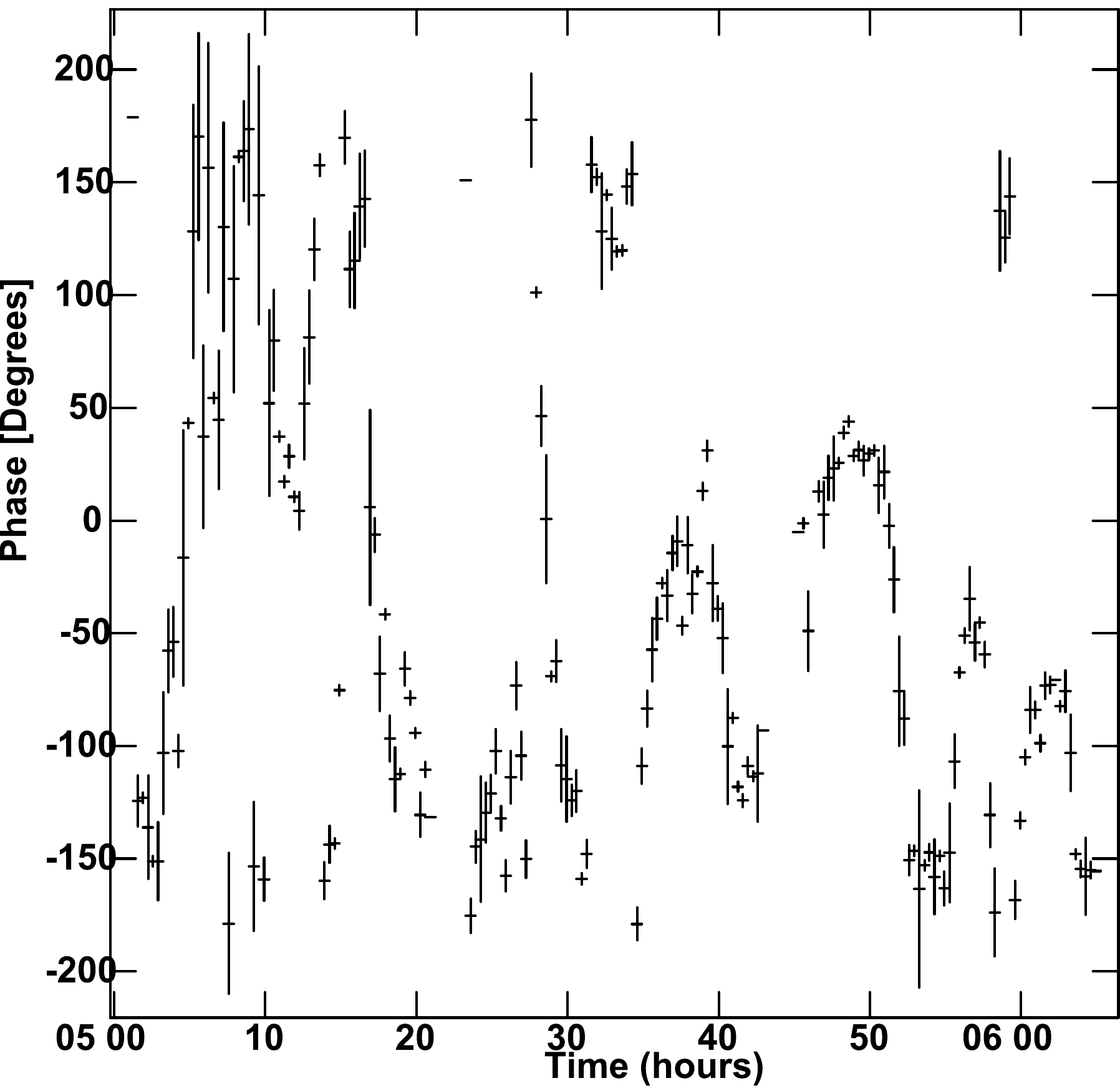}

    \end{tabular}
    
    \caption{The 10 s integration of the source 4C 55.28 in Stokes I in flux vs. time (top) and phase vs. time (bottom) observed May 28\textsuperscript{th}, 2018.  The increasing flux density throughout the observation is a result of a similar trend in the calibration beam, but the distinct offset from 0 Jy of the signal makes us confident in the detection.}
    \label{fig:Method_Check}
\end{figure}

  
We spent a total of 134.88 hours on source across the five stars, observing them at 68 and 78 MHz with a 9.8 MHz bandwidth for each of two frequency tunings (a total bandwidth of 19.6 MHz between 63.7 and 83.3 MHz). Following the observations, we used an LSL-based software correlator (Dowell et al., 2012) to correlate the data and provide an initial flag table. After correlation, we used AIPS in order to further flag and reduce the data. Signals in the calibration beam that appeared to be RFI were flagged and these flag tables were copied over to the source beam.  There was further clean up in the source beam for signals that happened in single frequency channels and were short in time.  Because flares are typically between a few seconds and several minutes, we looked at 1, 10, and 100 second integrations and averaged over the 20 MHz observing bandwidth to increase sensitivity for initial data reduction. Using this method, we were able to make a detection of the strong continuum source 4C 55.28, a 4.5 Jy source (see Figure \ref{fig:Method_Check}), placing a lower limit on our detection capabilites.  An extra condition that was employed when deciding whether a flare was detected was the signal's presence in Stokes V due to the emission regularly being circularly polarized for brief flares (Lang \& Willson 1986; Bastian \& Bookbinder 1987; Osten \textit{et al.} 2005; Lynch \textit{et al.} 2017).
\begin{table*}
    \small
    \centering
    \begin{tabular}{r|r|l|r|l|r|l|c}
   \hline \hline
     \multicolumn{1}{r}{\textbf{Star}} &\multicolumn{1}{c}{\textbf{Phase Cal}} & \multicolumn{1}{c}{\textbf{Flux}}& \multicolumn{1}{c}{\textbf{Flux Cal}} &\multicolumn{1}{c}{\textbf{Flux}}& \multicolumn{1}{c}{\textbf{Check Source}} & \multicolumn{1}{c}{\textbf{Flux}}&
     \multicolumn{1}{r}{\textbf{Time on Source [hrs]}} \\ \hline
     AD Leo & 3C 241 & 17.8 & 3C 295 & 125 & 3C 286 & 28.0 & 45.2\\
     Wolf 424 & 4C 09.41& 22.6 & 3C 295 & 125 & 4C 48.38 & 13.9 & 6.90\\
     EQ Peg & 4C 17.94 & 12.5 & 3C 48 & 67.6 & 3C 41 & 20.4 & 32.2\\
     EV Lac & 4C 46.47 & 11.2 & 3C 418 & 23.7 & 3C 415.2 & 21.6 & 29.9\\
     UV Ceti & PKS 0139-16 & 17.1 & 3C 48 & 67.6 & 3C 41 & 20.4 & 20.7\\ \hline \hline 
     
    \end{tabular}
    \caption{Information on the flux and phase calibrator and check source used for the observations.  On the left in each column is the name of the source and on the right is the flux density in Jy at 74 MHz. The locations and flux densities of the calibrators and check sources were provided by the NASA/IPAC Extragalactic Database (NED).}
    \label{tab:obs_info}
\end{table*}

\section{Results}\label{sec:Results}

\subsection{AD Leo}
\begin{figure*}
    \centering
    
    \includegraphics[width=0.68\textwidth]{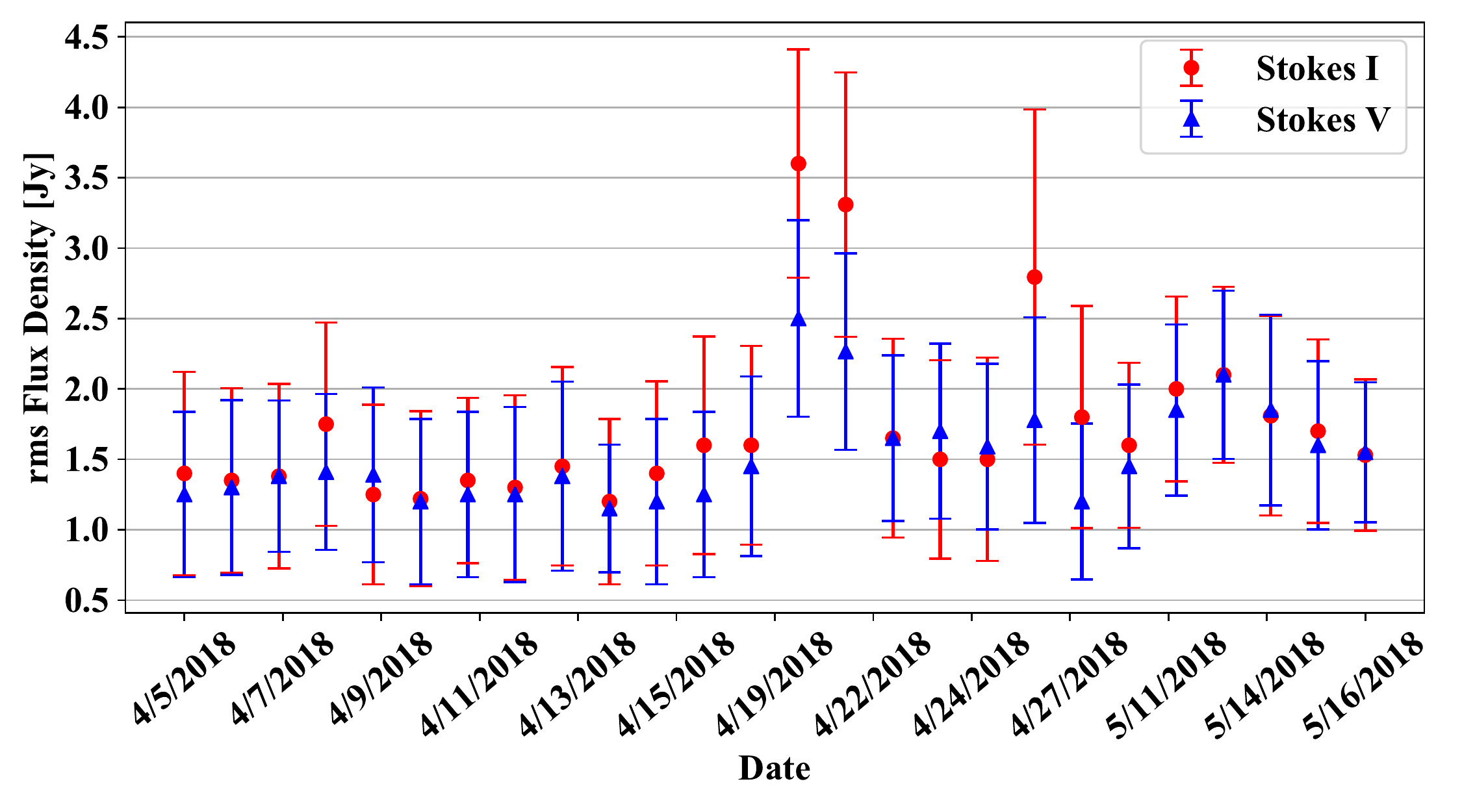}
    \caption{The rms and maximum values of 10 second integrations for each AD Leo observation; the marker is at the peak value and the rms is represented by the error bars.}
    \label{fig:AD_Leo_rms}
\end{figure*}

AD Leo was observed for 45.2 hours across 26 days from April 5\textsuperscript{th} to May 17\textsuperscript{th} in 2018. From Figure \ref{fig:AD_Leo_rms}, the observations on April 21\textsuperscript{st} and 22\textsuperscript{nd} appear to have significant detections in both Stokes I and V in the ten second integrations, however, these peak values were not coincident, with their separations in time being as much as 4.5 minutes apart.  These cases also showed no evidence of time evolution; rather than there being a rise and decay in the flux, these peak values corresponded to single instances, making it extremely unlikely they were true flare events. \par

\subsection{Wolf 424}
Wolf 424 is the flare star we have the least amount of information on as only three of the 14 scheduled observations were successful due to high ambient temperatures at one of the LWA sites which prevented good operation. For these three observations, there is no evidence that a flare occurred.  Because so little time was spent on this object and the lack of detections, we cannot reasonably make any strong conclusions about the flaring properties or activity of Wolf 424.

\subsection{EV Lac}
\begin{figure*}%[hhhhhh!!]
    \centering
    \includegraphics[width=0.68\textwidth]{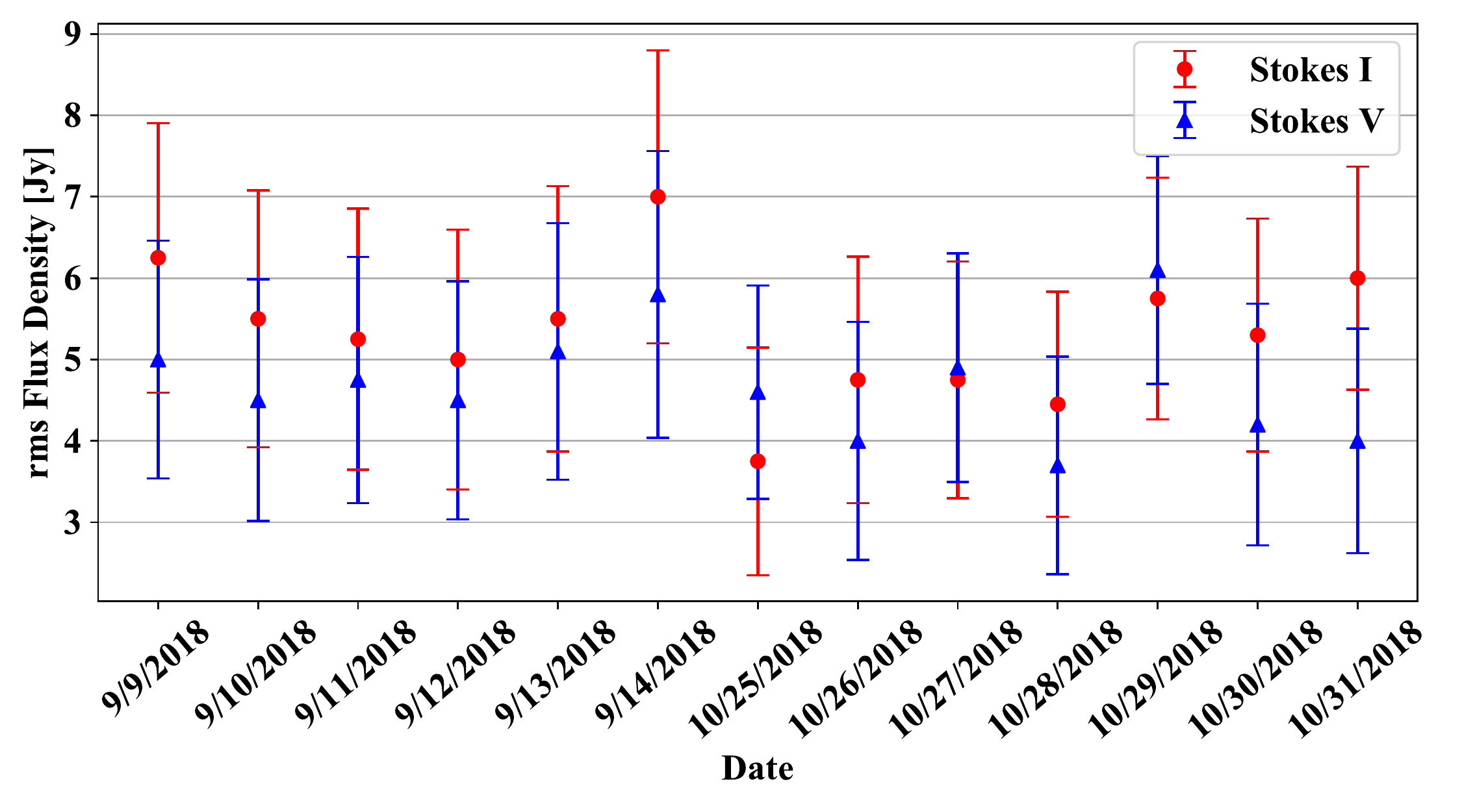}
    \caption{The rms and maximum values of 10 second integrations for each EV Lac observation.}
    \label{fig:Lac_rms}
\end{figure*}

EV Lac was observed for 29.9 hours across 13 days in early September and late October in 2018.  Although in Figure \ref{fig:Lac_rms} it would seem that the peak value in the observations of EV Lac were consistently significantly greater than the root mean square (rms), the flux density versus time plots for each observation show the flux density for both the star and the phase calibrator decreasing steadily throughout the observation. We attributed this trend to the proximity of EV Lac and 4C 46.47 to the Galactic plane during the observations, making it impossible to make conclusions about any possibly occuring flares at these low levels.

\subsection{UV Ceti}
\begin{figure*}%[hhhhhh!!]
    \centering
    \includegraphics[width=0.68\textwidth]{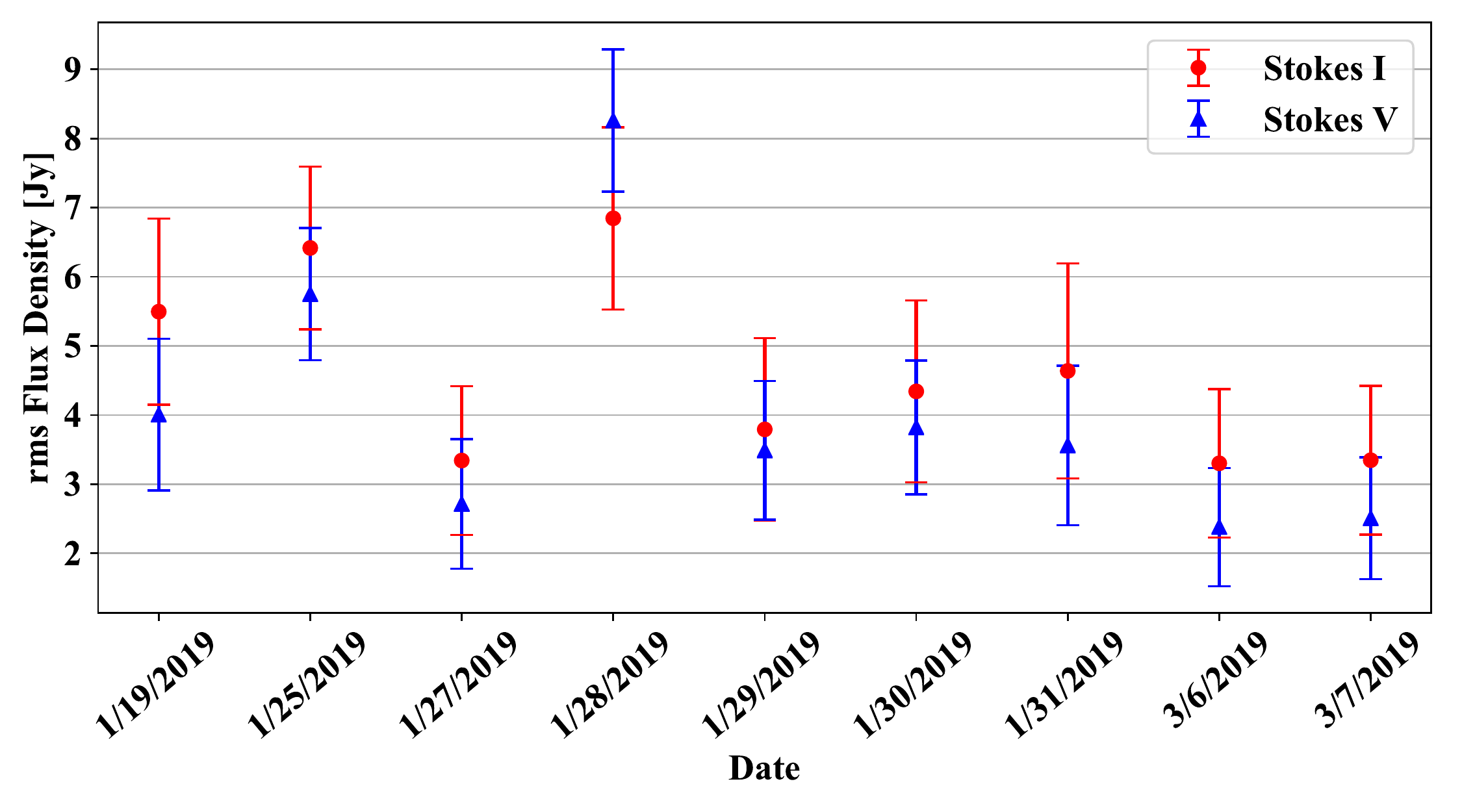}
    \caption{The rms and maximum values of 10 second integrations for each UV Ceti observation.}
    \label{fig:Ceti_rms}
\end{figure*}

UV Ceti was observed for 20.7 hours across 9 days in late January and early March in 2019. In each observation, there was an apparent increase in the flux density in both Stokes I and V in the last hour of the $\sim$3 hour observation runs.  This signal was consistently present in the 10 s integrations and often in the 100 s integrations as well.  However, this signal was found to be due to a source passing through the fan beam that was centered on UV Ceti towards the end of the run.  Because of the interference of the source, none of the peak values for these observations (see Figure \ref{fig:Ceti_rms}) can be attributed to flaring from the star and no other times during the observations showed evidence of flaring activity.\par

\subsection{EQ Peg}

\begin{figure}
    \centering
    \begin{tabular}{cc}
        \includegraphics[width= 0.43\textwidth]{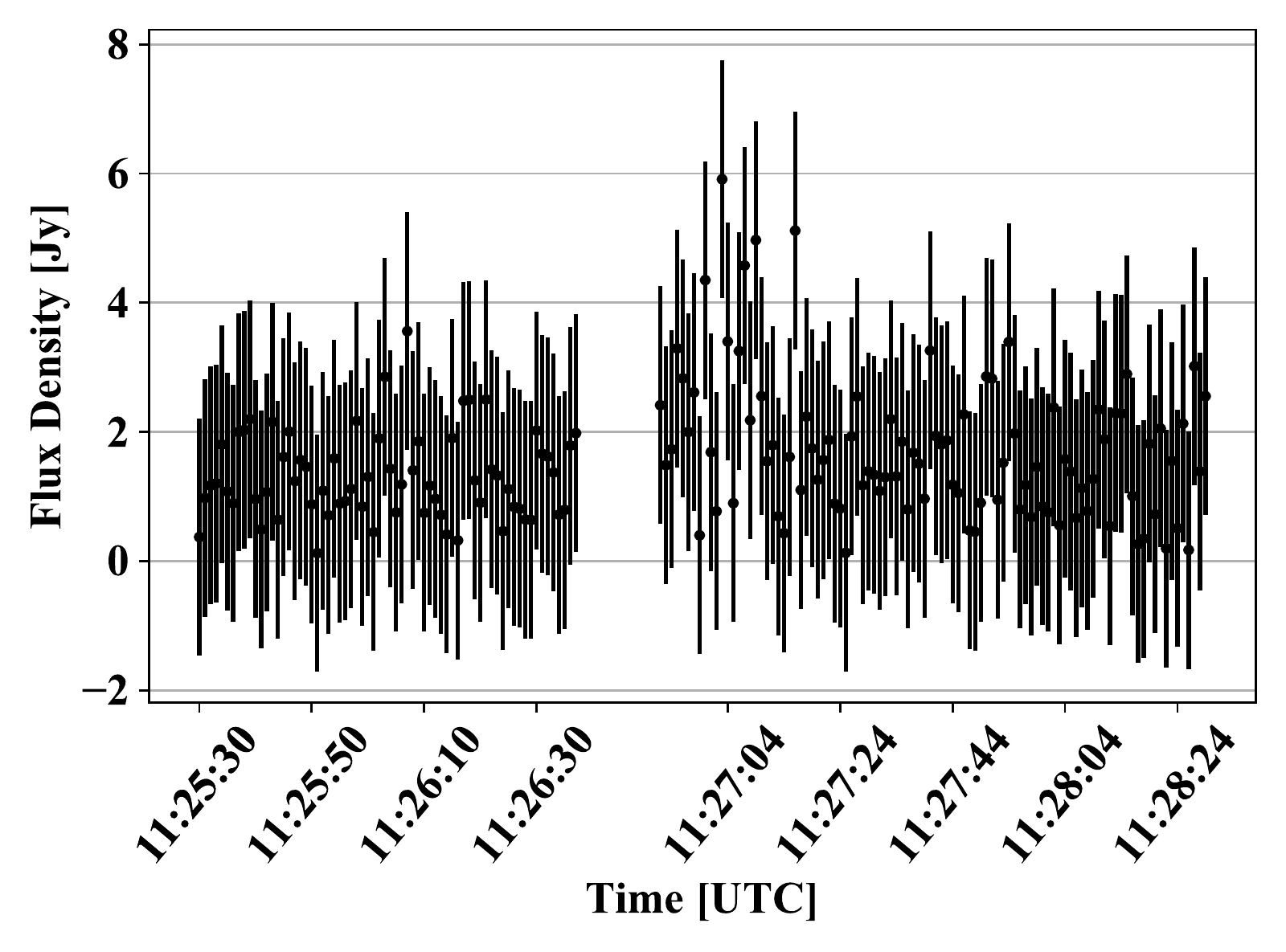} \\
         \includegraphics[width= 0.43\textwidth]{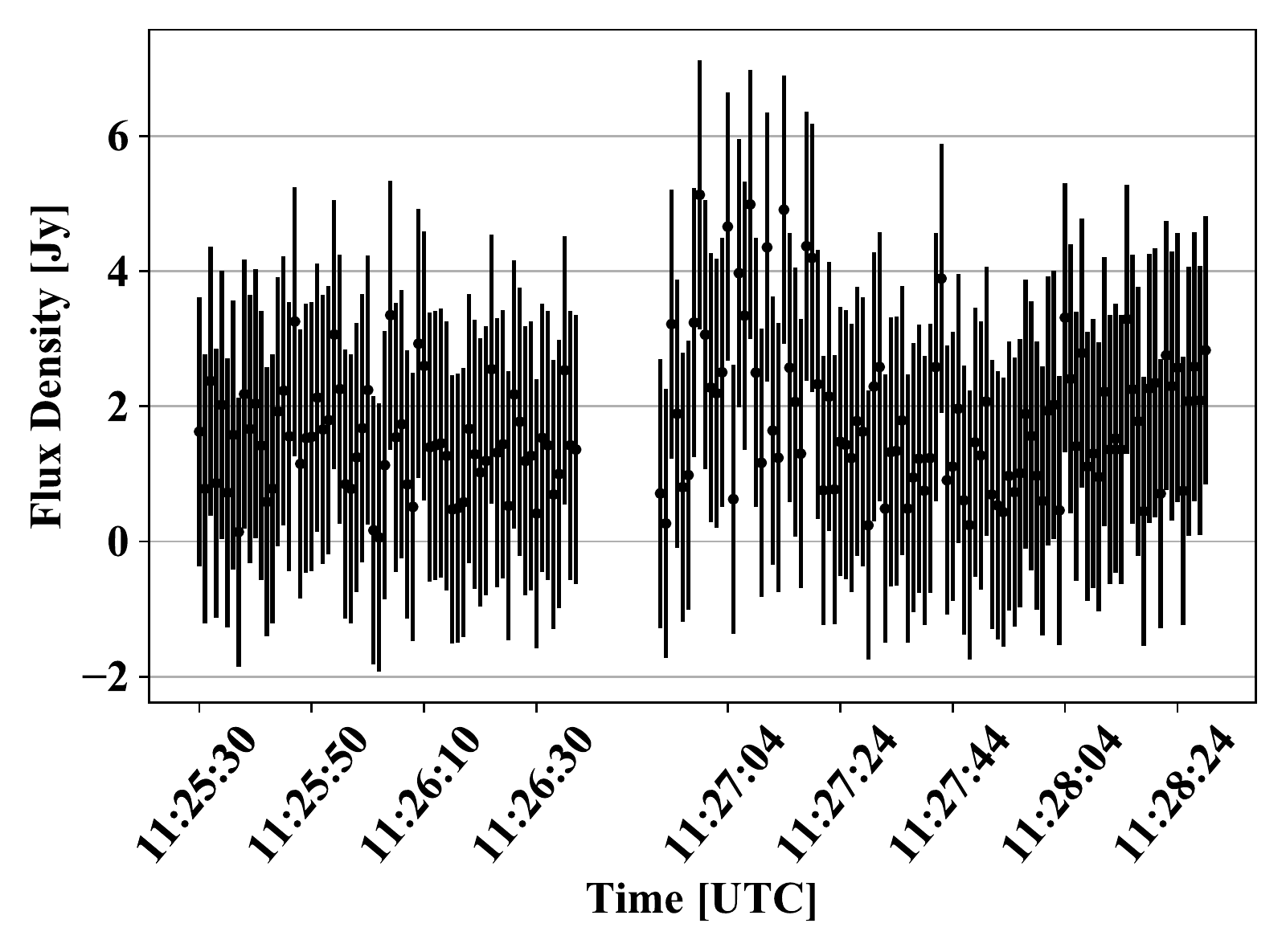} 
    \end{tabular}
    \caption{A slice from the 1 s integrations of EQ Peg on July 20\textsuperscript{th}, 2018, centered on the flare candidate at around UTC 11:27:05 in Stokes I (top) and Stokes V (bottom). The total height of the error bars is twice the rms of the observation.}
    \label{fig:EQPeg_Flare}
\end{figure}

\begin{figure*}
    \centering
    \includegraphics[width=0.68\textwidth]{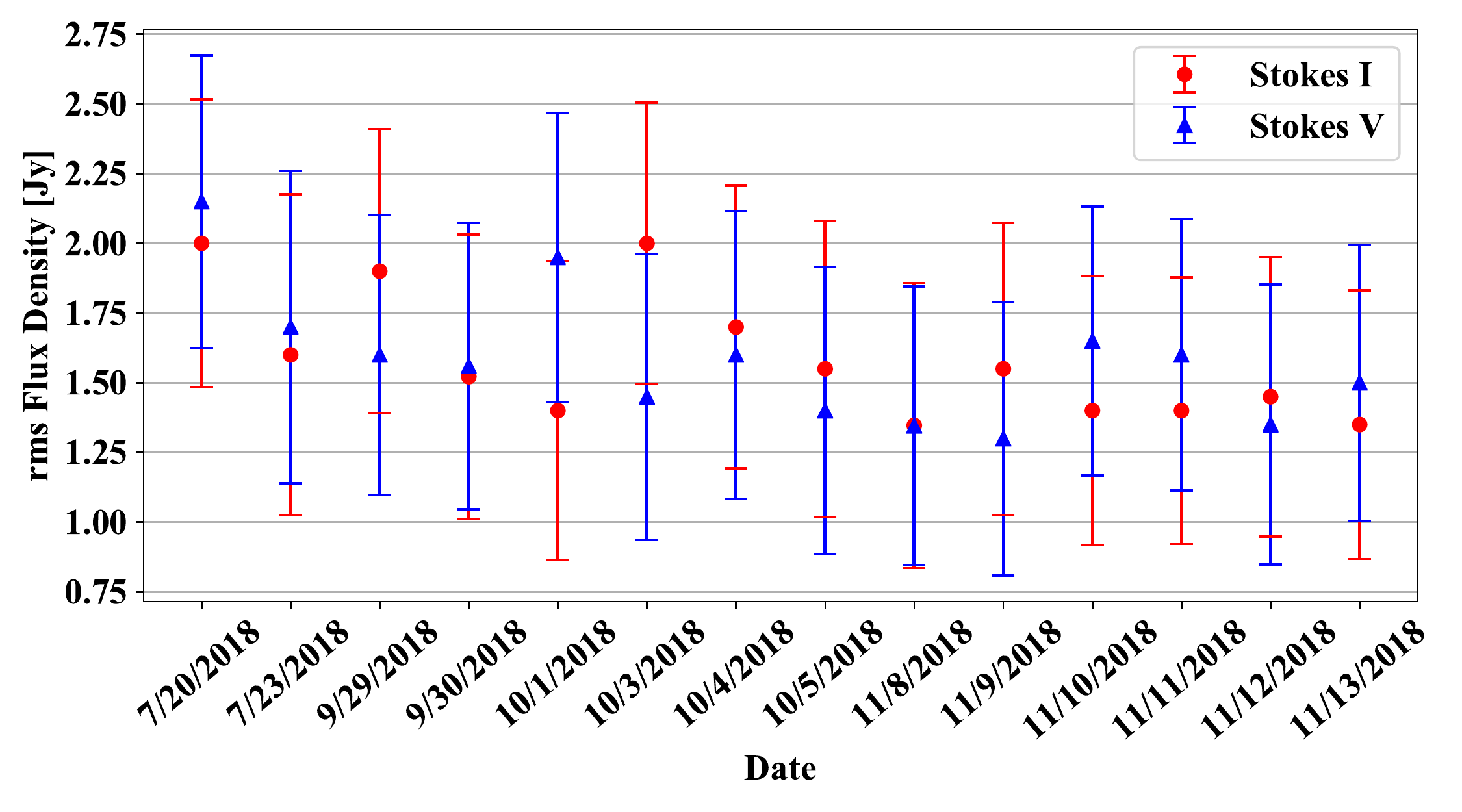}
    \caption{The rms and maximum values of 10 second integrations for each EQ Peg observation.}
    \label{fig:EQ_Peg_rms}
\end{figure*}

EQ Peg was observed for a total of 32.2 hours across 14 days from July 20\textsuperscript{th} to November 13\textsuperscript{th} in 2018. There is strong evidence of a flare  on July 20\textsuperscript{th} at 11:27:10 UTC, shown in Figure \ref{fig:EQPeg_Flare}.  While the flare reached about 5.91 Jy for Stokes V and 5.13 Jy in Stokes V for the 1 second integrations, it was only 2 Jy in Stokes I and 2.15 Jy in Stokes V during the ten second integrations (Figure \ref{fig:EQ_Peg_rms}) and was virtually absent during the 100 second integrations.  This could be because of phases not adding coherently due to ionospheric effects.

\section{Discussion}
Our best flare candidate was from EQ Peg on July 20\textsuperscript{th}, 2018.  Although the signal for the flare was significantly diminished in the 10 and 100 s integrations, the flare lasted for $\sim$30 seconds in the 1 s integrations. The absence of the flare during longer integrations, particularly in the 100 s integration, is likely the result of a rapidly varying phase due to a disturbed ionosphere. When averaged over the entire 19.6 MHz bandwidth, the Stokes I peak value of the flare was 5.91 Jy with an rms of 1.84 Jy for 1 s integrations.  In Stokes V, the peak value was 5.13 Jy with an rms of 1.99 Jy. The fact that the flux density in Stokes V was so close to the value of the flare in Stokes I suggests that this signal was significantly circularly polarized, further supporting that this may have been a true flare event, making this the lowest frequency detection of EQ Peg to date.\par

When inspecting the calibration corrected data prior to frequency averaging, no particularly strong channel was found in the cross power spectrum in the time range of the flare, leading us to believe this was a broadband flare.  Because the degree of our spectral analysis is so sensitivity-limited, this emission cannot be differentiated between plasma emission or multiple ECM sites via observational results. When considering the magnetic and plasma environment of EQ Peg, we use magnetic field strengths of 480 G for EQ Peg A and 450 G for EQ Peg B (Donati \textit{et al.}, 2008) in the equation for cyclotron frequency in MHz (equation \ref{eq:cyc_freq}), where B is the magnetic field in Gauss, gives cyclotron frequencies of 1.34 and 1.26 GHz for the A and B components respectively.

\begin{equation}
    \nu_{cyc} = \frac{eB}{2\pi m_ec} \approx 2.8B,
    \label{eq:cyc_freq}
\end{equation}

To calculate the plasma frequency, we use equation \ref{eq:plasma_freq} where $m$ is the effective mass of the electron, $q$ is the electric charge, and $n$ is the electron number density. Using $3\times10^{11}$ cm$^{-3}$ and $4\times10^{10}$ cm$^{-3}$ as the upper and lower limits on the  number density of EQ Peg's corona (Liefke, Ness, Schmitt \& Maggio 2008) gives plasma frequencies of 4.92 and 1.80 GHz for the upper and lower plasma density estimates respectively. This puts the fraction $\nu_{cyc}/\nu_p$ between 0.26 and 0.75.  For fundamental ECM emission to escape,  $\nu_{cyc}/\nu_p \gtrsim 1$, although harmonic emission is possible for $2\nu_{cyc}/\nu_p > 1$ and thus is a possibility. Because either emission process should be possible even for the globally averaged magnetic field strength and number density, as well as the fact that local conditions may exist in the field and plasma not accounted for in the average values that would make even fundamental ECM possible, we cannot rule out ECM or plasma emission as being responsible for this flare in this way.
\begin{equation}
    \nu_p = \frac{1}{2\pi}\sqrt{\frac{4\pi nq^2}{m}} \approx 9000\sqrt{n},
    \label{eq:plasma_freq}
\end{equation}

\par
Using equation \ref{eq:bright_temp} where $T\textsubscript{B}$ is the brightness temperature, $S_{\nu}$ is the flux density, $D$ is the distance to the source, and $r_s$ is the radius of the emission source, we find $T_{B} = \ 1.75 \times 10^{16}(r/r_*)^{-2}$ K assuming a stellar radius $r_* =$ 0.23\(R_\odot\) (White, Jackson, \& Kundu 1989), a temperature achievable for both ECM and fundamental plasma emission. Although the duration of this flare ($\sim30$ s) is reminiscent of the duration of the broadband, plasma emission related type III solar bursts, the polarization of type III bursts is little to none at low frequencies (Reid \& Ratcliffe, 2014). The likely ECM driven type IV bursts are also broadband and highly circularly polarized.  However, this kind of burst is often longer in duration ($> 10$ minutes) (Liu \textit{et al.}, 2018), although this duration difference could possibly be accounted for by a difference in beaming geometry between EQ Peg and the Sun should ECM be the responsible process. It is also possible that the flare lasted longer, but we were only able to detect it during the time it was brightest. Without a good solar analogy with which to compare this flare event to, we are still unable to determine the emission process beyond it likely being fundamental plasma or ECM emission.\par 
\begin{equation}
    T_B = \frac{S_{\nu}}{2k}\frac{c^2}{\nu^2}\frac{D^2}{\pi r_s^2}
    \label{eq:bright_temp}
\end{equation}

The possible processes' dependence on the environment of the star, and this environment's variability, could explain why there is such variability in what frequencies EQ Peg-- as well as other flare stars bright in the radio-- can be seen at as well as in what processes are predicted to be responsible.  EQ Peg has been successfully detected at frequencies as high as 1.4 GHz (Kundu, Pallavicini, White, \& Jackson 1988; Jackson, Kundu, \& White 1989) and as low as 320 MHz (Crosley \& Osten 2018), however, observations conducted by Arecibo at 430 MHz show no evidence of flaring activity (Spangler \& Shawhan 1976). Similar results are found for AD Leo; coordinated observations of AD Leo with the Clark Lake Radio Observatory and VLA had only two brief detections at 110.6 MHz but found it was consistently flaring in the time between those two detections at 1.4 GHz (Jackson, Kundu \& Kassim, 1990; White, Kundu, \& Jackson, 1986; Kundu \textit{et al.}, 1988).  However, using the UTR-2 array, Boiko \textit{et al.} (2012) made as many as 30 detections on a single day of observation with the flux density of the flares ranging between 4 and 300 Jy at frequencies between 16.5 and 33 MHz.  Although instrumental limitations or error could very well account for this, it is also not unlikely the variability of the environment of young stars with strong magnetic fields would drastically affect the emission frequency for processes which so heavily depend on the plasma density.

\section{Summary}
Many stars have been shown to exhibit flaring properties similar to those presented by the Sun across many parts of the electromagnetic spectrum.  However, results below 1 GHz have been largely inconclusive with observations showing huge variety in the frequency of the events, the wavelength at which they occur, the drift rate, flux density, and size of the area which the signal must be originating from.  The goal of the observations presented here was to provide further restrictions for these properties for low frequencies.  Although only one detection was made across the 130 hours spent on five flare stars, the properties of the flare from EQ Peg are consistent with characteristics of either ECM or fundamental plasma emission. This flare was broadband as well as had a brief duration-- suggesting a similar emission process as type III bursts solar flares-- and was strongly circularly polarized like type IV bursts. Without the possibility of better spectral analysis, a better understanding of the plasma density and magnetic field topologies, or an all-encompassing solar burst analogy, we cannot adequately discern which of the two processes was responsible.  Given the dependence of these processes on the coronal properties of the star, and that these properties are subject to significant variability for stars with strong, variable magnetic fields, it seems likely that our low duty cycle was heavily affected by the nature of the emission process.\par 
Although the nature of the possible emission processes was likely a significant contributor to the difficulty in detecting flares, the sensitivity of the two-element LWA may very well be an issue, especially in the case of AD Leo which received the most observation time and for which we still made no detections, despite having been observed at similar frequencies by others. Some obstacles encountered for observations of the other flare stars were the LWA shutting down to avoid overheating and interference from other sources in the fan beam. \par 
Ideally, coordinated observations at a variety of frequencies should be conducted for objects that did not experience confusion from other sources passing through the fan beam (AD Leo, Wolf 424, and EQ Peg) in order to better discern whether the low duty cycle is due to the sensitivity or the frequency range-- and consequently the emission process.  For objects which experienced significant confusion (EV Lac and UV Ceti), we need to choose observation times which better exclude the interfering sources without sacrificing too much of the time which the source spends close to its transit height. Despite only confidently detecting one flare, this flare provided the lowest frequency detection of EQ Peg to date and confirms the LWA as a viable instrument in stellar research at frequencies below 100 MHz.  The sensitivity of the observations can be improved with the implementation of the LWA Swarm concept (Dowell \textit{et al.}, 2018) which would allow for imaging the fields and consequently self-calibration to remove ionospheric effects and thus allow longer integrations.\par

\section*{Acknowledgements}

Construction of the LWA has been supported by the Office of Naval Research under Contract N00014-07-C-0147 and by the AFOSR. Support for operations and continuing development of the LWA1 is provided by the Air Force Research Laboratory and the National Science Foundation under grants AST-1835400 and AGS1708855. We would also like to thank Stephen White and the reviewer for their input.

\bibliographystyle{mnras}
\nocite{*}
\bibliography{flare_stars}
\label{lastpage}
\end{document}